\documentclass[aps,prl,twocolumn,superscriptaddress,showpacs,floatfix]{revtex4}

\usepackage{bm}
\usepackage{amsmath}
\usepackage{amssymb}
\usepackage{graphicx}
\usepackage{color}
\usepackage{soul,xcolor}
\setstcolor{blue}
\usepackage{ulem}
\usepackage{cancel}

\def\sig{{\mbox{\boldmath{$\sigma$}}}}

\makeatletter
\@ifundefined{textcolor}{}
{%
 \definecolor{BLACK}{gray}{0}
 \definecolor{WHITE}{gray}{1}
 \definecolor{RED}{rgb}{1,0,0}
 \definecolor{GREEN}{rgb}{0,1,0}
 \definecolor{BLUE}{rgb}{0,0,1}
 \definecolor{CYAN}{cmyk}{1,0,0,0}
 \definecolor{MAGENTA}{cmyk}{0,1,0,0}
 \definecolor{YELLOW}{cmyk}{0,0,1,0}
\definecolor{ORANGE}{rgb}{1,0,1} }

\@ifundefined{definecolor}
 {\@ifundefined{definecolor}
 {\@ifundefined{definecolor}
 {\@ifundefined{definecolor}
 {\@ifundefined{definecolor}
 {\usepackage{color}}{}
}{}
}{}
}{}
}{}

\usepackage{epsfig}

\def\nab{{\mbox{\boldmath{$\nabla$}}}}
\def\sig{{\mbox{\boldmath{$\sigma$}}}}

\def\b0{{\bf{0}}}

\begin{document}


\title{Electric and Magnetic  Gating of Rashba-Active  Weak Links}

\author{A. Aharony}
\email{aaharony@bgu.ac.il}
\affiliation{Raymond and Beverly Sackler School of Physics and Astronomy, Tel Aviv University, Tel Aviv 69978, Israel}
\affiliation{Physics Department, Ben Gurion University, Beer Sheva 84105, Israel}
\affiliation{IBS Center for Theoretical Physics of Complex Systems,
34051 Daejeon, South Korea}

\author{O. Entin-Wohlman}
\affiliation{Raymond and Beverly Sackler School of Physics and Astronomy, Tel Aviv University, Tel Aviv 69978, Israel}
\affiliation{Physics Department, Ben Gurion University, Beer Sheva 84105, Israel}
\affiliation{IBS Center for Theoretical Physics of Complex Systems,
34051 Daejeon, South Korea}

\author{M. Jonson}
\affiliation{Department of Physics, University of Gothenburg, SE-412
96 G{\" o}teborg, Sweden}
\affiliation{IBS Center for Theoretical Physics of Complex Systems,
34051 Daejeon, South Korea}

\author{R. I. Shekhter}
\affiliation{Department of Physics, University of Gothenburg, SE-412
96 G{\" o}teborg, Sweden}
\affiliation{IBS Center for Theoretical Physics of Complex Systems,
34051 Daejeon, South Korea}

\date{\today}

\begin{abstract}

In a one-dimensional weak-link wire the    spin-orbit interaction (SOI) alone cannot generate a nonzero spin current.
We show that a Zeeman field acting in the wire in conjunction with the Rashba SOI there does yield such a current, whose magnitude and direction depend on the direction of the field. When this field is not parallel to the effective field due to the SOI, both the charge and the spin currents oscillate with the length of the wire. Measuring the oscillating anisotropic magnetoresistance can thus yield information on the SOI strength.  These features  are tuned by applying  a magnetic and/or an electric field, with possible applications to spintronics.
 \end{abstract}


\maketitle

Spintronics takes advantage of the electronic spins in designing a variety of applications, including   giant magnetoresistance sensing, quantum computing,  and quantum-information processing \cite{wolf,zutic}.  A promising approach for the latter  exploits mobile qubits, which carry the quantum information via the spin polarization of the moving electrons. The spins of mobile electrons can be manipulated  by  the spin-orbit interaction (SOI), which  causes  the spin of an electron moving through a spin-orbit active material (e.g.,     semiconductor heterostructures \cite{Kohda})  to rotate around an effective magnetic field that depends on the momentum \cite{winkler,manchon}. In the particular case of the Rashba SOI \cite{rashba}, both the rotation axis and the amount of rotation can be tuned by  gate voltages \cite{Nitta,Sato,Beukman,comDres}.
Research in  this  direction was  enhanced following the proposal by 
Datta and Das  \cite{datta},
of a spin field-effect transistor  based on magnetic leads. It is still a
challenge to achieve polarized mobile electronic spins avoiding the use of  ferromagnetic leads.

In the simplest device, electrons move between two large electronic reservoirs, via a mesoscopic region. When the region is spin-orbit active, the single-channel transmission  is described by a $2\times 2$ matrix in spin Hilbert space.  Since this matrix is proportional to the unit matrix when time-reversal symmetry is obeyed  \cite{bardarson},  spin splitting cannot be achieved with SOI alone. Time-reversal symmetry is broken by applying  a magnetic field. Indeed,  several  proposed devices  utilize an orbital Aharonov-Bohm magnetic flux, which penetrates loops of interferometers to achieve spin splitting \cite{lyanda,us,Saarikoski}, via  
the  interference of the spinor wave functions in the two branches of the loop. 

\begin{figure}[htp]
\includegraphics[width=5cm]{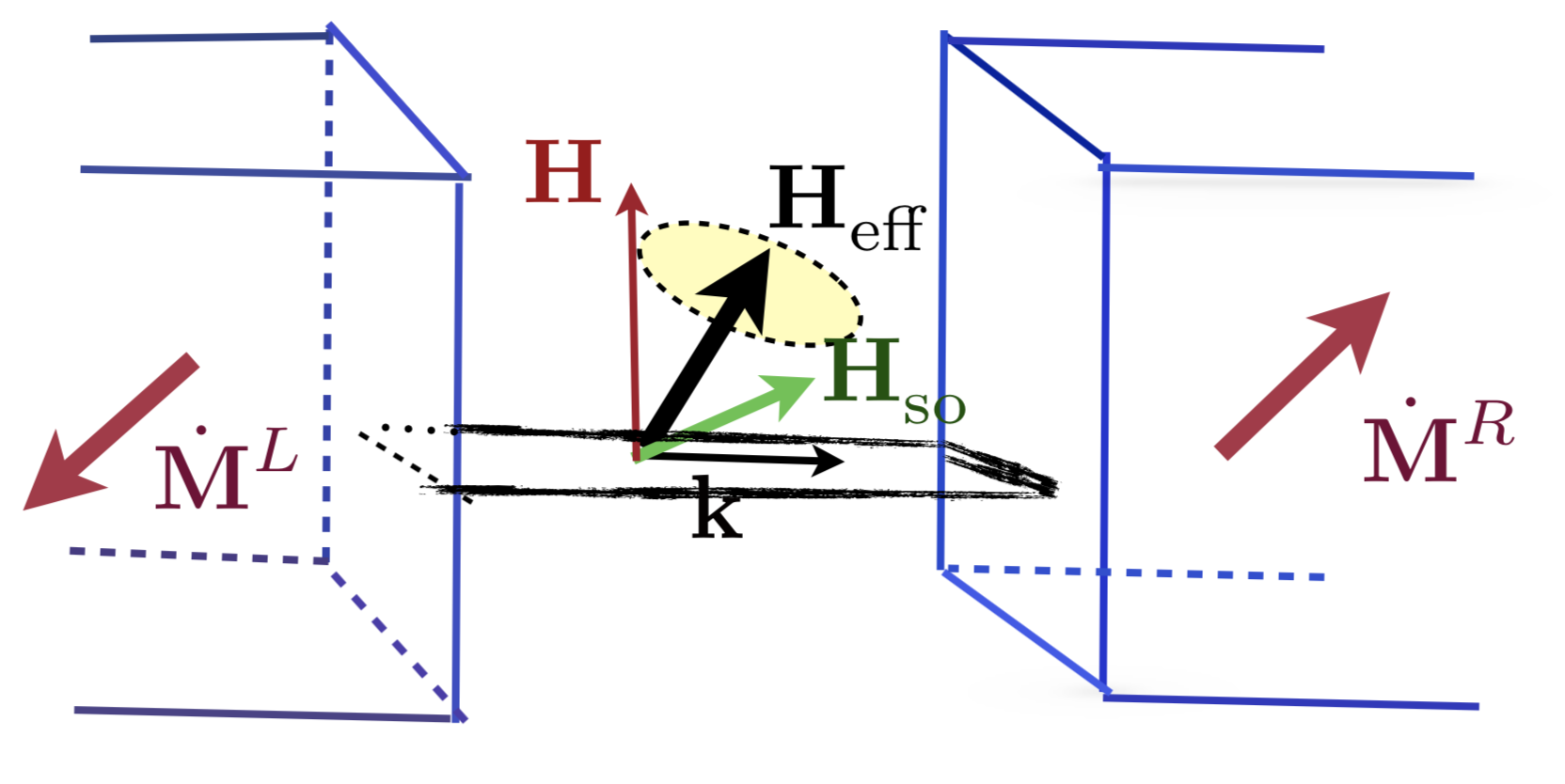}
\caption{(Color online.) A spin-orbit active weak-link wire connecting  two reservoirs. The momentum of the electron is ${\bf k}$, the external magnetic field is ${\bf H}$ and the effective field due to the SOI is ${\bf H}^{}_{\rm so}$. The net effective field (around which the spins rotate) is $ {\bf H}^{}_{\rm eff}$. The arrows in the reservoirs show  the  magnetization rates there, $\dot{\bf M}^{L}$ and $\dot{\bf M}^{R}=-\dot{\bf M}^{L}_{}$, generated due to the joint effect of ${\bf H}_{\rm so}$ and ${\bf H}$.
 }
\label{f1}
\end{figure}

Here we  analyze an even simpler geometry: the two reservoirs are connected by a {\it single} (weak link) spin-orbit active wire, but we do take into account the Zeeman energy gained from an external  magnetic  field acting on the whole wire (see Fig. \ref{f1}). Due to 
this field, both the charge and the spin conductances of the device are found to exhibit oscillations with the length of the wire.  These oscillations, as well as
the associated magnetoconductance anisotropy, can be used to identify the strength of the SOI; remarkably, they can be tuned by applying electric and/or magnetic fields. Although earlier papers \cite{privman,goldhaber,gudmundsson} considered the band structure and the gate-voltage dependence of the conductance of such wires, they did not discuss these interesting phenomena.  
 
The SOI and the Zeeman field split the spinor wave function in the wire into two waves, with different wave vectors and  with different spin  polarizations. In the presence of both the external magnetic field,  ${\bf H}$,  and the effective magnetic field due to the SOI,  ${\bf H}^{}_{\rm
 so}$,  each of these   accumulates its own phase along the link, and they interfere to generate a 2$\times$2 matrix for the tunneling amplitude, which contains a mixture of the two polarizations.
Without the external field, this matrix  is unitary, causing only a rotation of  the spin polarizations around ${\bf H}^{}_{\rm
 so}$ by an angle which depends on the SOI strength and the wire length \cite{Oreg}. The resulting transmission matrix is proportional to the unit matrix, and there is no net polarization. In contrast,  a nonzero ${\bf H}$ generates a non-unitary tunneling amplitude.  Then, a bias voltage between the reservoirs induces   particle {\it and} spin currents. 
Thus, the non-unitarity of the transmission matrix results in a net spin magnetization  in the reservoirs. This,   as well as the magnetoconductance of the weak link, do not depend on the length of the wire {\it unless} the magnetic field is not along  ${\bf H}^{}_{\rm
 so}$. In particular, when ${\bf H}$  is perpendicular to
 ${\bf H}^{}_{\rm
 so}$, both properties  exhibit
distinct oscillations with this length, on length scales related to the SOI and the field strengths. As such, they    are both tunable by  electric {\it and}  magnetic fields. The  magnetization generated in the reservoirs has components along ${\bf H}$ and  ${\bf H}^{}_{\rm
 so}$, and - surprisingly - also perpendicular to both of these directions. Remarkably, these 
 effects are of the order of the ratio of the Zeeman energy to the spin-orbit energy \cite{comAB}. 

The described interference can be compared to the Aharonov-Bohm effect, where the electron wave function acquires different extra phases from the motion of an electric charge along different paths in an external magnetic field. Here the extra phases - which we may call Aharonov-Casher  \cite{AC} phases for electrons - are due to the motion of a magnetic moment in an electric field. Our device can therefore be viewed as an Aharonov-Casher interferometer. Unlike the Aharonov-Bohm one, here the phase difference between two ``channels" can appear even though the electrons move along the same spatial trajectory between two singly-connected leads.


Adopting units in which $\hbar=1$ and using the linear  Rashba SOI,  
the Hamiltonian of the system  is 
\begin{align}
{\cal H}={\cal H}^{}_{\rm link}+{\cal H}^{}_{\rm leads}+{\cal H}^{}_{\rm tun}\ .
\end{align}
Here, the Hamiltonian in the weak link is
\begin{align}
{\cal H}^{}_{\rm link}=-\frac{1}{2m^{\ast}_{}}\nab^2_{}
-i\frac{\widetilde{k}^{}_{\rm so}}{m^{\ast}_{}} \hat{\bf n}\cdot(\sig\times\nab)-{\bf H}\cdot\sig\ ,
\label{ham}
\end{align}
where $\sig$ is the vector of Pauli matrices,   $m^{\ast}$ is the effective mass, 
and $\hat{\bf n}$ is a unit vector along the electric field which causes the SOI.  The net strength of this interaction (in momentum units \cite{comRashba})  is denoted $\widetilde{k}_{
\rm so}$;   the notation $k_{\rm so}$  is used below for the ``full" coupling.   The ``magnetic field" ${\bf H}$ contains
the factor  $(g/2)\mu_{\rm B}$, i.e., 
the $g-$factor and 
the Bohr magneton,   
and therefore has units of energy.  The Hamiltonian of the leads is ${\cal H}^{}_{\rm leads}=
\sum_{\alpha=L,R}{\cal H}^{\alpha}_{\rm lead}$, with
\begin{align}
{\cal H}^{\alpha=L(R)}_{\rm lead}=\sum_{{\bf k} ({\bf p}),\sigma}\epsilon^{}_{ k(p)}c^{\dagger}_{{\bf k}({\bf p})\sigma}c^{}_{{\bf k}({\bf p})\sigma}
\ ,
\label{Hl}
\end{align}
where $c^{\dagger}_{{\bf k}\sigma}$ ($c^{}_{{\bf k}\sigma}$) creates (annihilates) an electron with momentum ${\bf k}$ and spin $\sigma$  (at an arbitrary quantization axis at this stage) in the left lead, with similar definitions in the right lead.  The tunneling between the leads and the weak link is described by
\begin{align}
{\cal H}^{}_{\rm tun}=
\sum_{{\bf k},{\bf p},\sigma,\sigma'}([V^{}_{{\bf k}{\bf p}}]^{}_{\sigma\sigma'}c^{\dagger}_{{\bf k}\sigma}c^{}_{{\bf p}\sigma'}+[V^{\ast}_{{\bf k}{\bf p}}]^{}_{\sigma\sigma'}c^{\dagger}_{{\bf p}\sigma'}c^{}_{{\bf k}\sigma})\ ,
\label{Htun}
\end{align}
where $[V^{}_{{\bf k}{\bf p}}]^{}_{\sigma\sigma'}$ is the tunneling amplitude from the state with momentum ${\bf p}$ and spin $\sigma'$ in the right lead to the state with momentum ${\bf k}$ and spin $\sigma$ in the left one. This amplitude, the key ingredient of our approach,  is proportional to the spin-dependent  propagator  connecting  the two states \cite{revRashba}.

Our calculation contains three steps. First,  ${\cal H}^{}_{\rm link}$ is used  to derive the propagator, i.e.,  the Green's function connecting a pair of electronic 
spin states  along the (one-dimensional) wire.   Second, the time derivatives of the particle number and the total spin  in the reservoirs is found to second-order 
in ${\cal H}^{}_{\rm tun}$.   For unpolarized leads  the Fermi distribution, e.g.,  in the left lead,  is
\begin{align}
f^{}_{L}(\epsilon^{}_{k})=1/[e^{\beta^{}_{}(\epsilon^{}_{k}-\mu^{}_{L})}+1]\ ,
\label{Fermi}
\end{align}
where $\beta^{}=(k^{}_BT)^{-1}$  is the inverse  temperature and $\mu^{}_L$ is  the chemical potential. The Fermi distribution in the right reservoir, $f^{}_R(\epsilon^{}_p)$, is defined similarly.
Finally, 
the transmission matrix and  the particle and spin currents are analyzed.
 We consider the case where the Fermi energy in the conducting wire,  $E^{}_{\rm F}$ ($E_{\rm F}$ is the common chemical potential of the reservoirs) exceeds significantly all other energies.
The opposite limit of an insulating wire ($E^{}_{\rm F}<0$  in our notation) is addressed in Ref. \onlinecite{Jonckheere};   an intriguing  interplay between $|E^{}_{\rm F}|$ and both the Zeeman and SOI energies is found to dominate the transport.

We first consider the propagator.
Assuming a plane-wave solution with a wave vector ${\bf k}$ directed along the one-dimensional wire, $\exp[i{\bf k}\cdot{\bf r}]=\exp[iks]$  ($s>0$ is the coordinate along the wire),
the effective magnetic field associated with the SOI is
\begin{align}
{\bf H}^{}_{\rm so}({\bf k})=(k k_{\rm so}/m^{\ast}_{})\hat{\bf h}^{}_{\rm so}\ ;
\quad \hat{{\bf h}}^{}_{\rm so}= (\hat{\bf n}\times\hat{\bf k})/\vert \hat{\bf n}\times\hat{\bf k} \vert\ ,
\label{Hso}
\end{align}
where $k_{\rm so} = \tilde{k}^{}_{\rm so} \vert \hat{\bf n}\times\hat{\bf k} \vert$ and $\hat{{\bf h}}^{}_{\rm so}$
is a unit vector along the direction of ${\bf H}_{\rm so}$ \cite{comRashba}. Then 
${\cal H}^{}_{\rm link}({\bf k})=k^{2}/(2m^{\ast}_{})-{\bf H}^{}_{\rm eff}({\bf k})\cdot\sig$,
with the net effective magnetic field
\begin{align}
{\bf H}^{}_{\rm eff}({\bf k})={\bf H}_{\rm so}({\bf k})+{\bf H}=\frac{k k_{\rm so}}{m^{\ast}_{}}\hat{\bf h}^{}_{\rm so}+{\bf H}\ .
\label{tildeH}
\end{align}
The (spin-dependent) propagator between two points along the link is a 2$\times$2 matrix in spin space \cite{revRashba,Shahbazyan}
\begin{align}
G(s;E)&=
\int dk e^{iks}\frac{E+i0^{+}_{}-\frac{k^{2}}{2m^{\ast}}-{\bf H}^{}_{\rm eff}({\bf k})\cdot\sig}{(E+i0^{+}_{}-\frac{k^{2}}{2m^{\ast}})^{2}-{H}^{2}_{\rm eff}({\bf k})}\ .
\label{Gn}
\end{align}
It is evaluated by  the Cauchy theorem \cite{sup}, with   
$E= k^{2}_{\rm F}/(2m^{\ast}_{})$, as the electrons in the link are at the Fermi energy.
The free-particle {propagator, $
G^{}_{0}(s;E)
=-i\pi m^{\ast}_{} \exp[ik^{}_{\rm F}s]/k^{}_{\rm F}
$,  is recovered when 
 ${\bf H}={\bf H}_{\rm so}=0$.

In the presence of the  effective magnetic field Eq. (\ref{tildeH}),  the external magnetic field ${\bf H}$ is decomposed  into components parallel ($H_{\parallel}$) and perpendicular ($H_{\perp}$)
 to $\hat{\bf h}^{}_{\rm so}$. The  poles of the integrand in Eq. (\ref{Gn}) are the  solutions of 
\begin{align}
(k^2-k_{\rm F}^2)^2
=(2m^{\ast}_{}H^{}_\perp)^2+(2k^{}_{\rm so}k+2m^{\ast}_{}H^{}_{\parallel})^2
\label{quad}
\end{align}
in the upper half of the complex $k-$plane \cite{compole}.  
Denoting these  by $k^{}_\pm$, with $k^2_\pm-k_{\rm F}^2=\pm2m^{\ast}_{}H^{}_{\rm eff}(k^{}_\pm)$, the propagator is
\begin{align}
&G(s;E)=G^{}_{0}(s;E)\exp[-ik^{}_{\rm F}s]\nonumber\\
&\times\big[e^{ik^{}_+s}A^{}_+(1+\hat{\bf q}^{}_+\cdot\sig)
+e^{ik^{}_-s}A^{}_-(1-\hat{\bf q}^{}_-\cdot\sig)\big]\ ,
\label{GG}
\end{align}
where the real coefficients $A^{}_\pm$ (i.e.,  the residues of the corresponding poles) and  the unit vectors [see Eq. (\ref{tildeH})]
\begin{align}
\hat{\bf q}^{}_\pm\equiv& \frac{{\bf H}^{}_{\rm eff}(k^{}_\pm)}{H^{}_{\rm eff}(k^{}_\pm)}
\label{qpm}
\end{align}
 depend on $k^{}_\pm$. The two terms in the square brackets of Eq. (\ref{GG}) correspond to waves with wave numbers $k^{}_+$ and $k^{}_-$ \cite{compole}. The corresponding tunneling amplitudes  contain the spin projection matrices  $(1\pm \hat{\bf q}^{}_\pm\cdot\sig)$, so that the transmitted electrons are fully polarized along $\hat{\bf q}^{}_+$ and $-\hat{\bf q}^{}_-$, respectively.

At zero magnetic field  
$\hat{\bf q}_+=\hat{\bf q}_-=\hat{\bf h}^{}_{\rm so}$,
and the propagator is proportional to a unitary matrix,
$G(s;E)\propto 
\exp[-ik^{}_{\rm so}s \hat{\bf h}^{}_{\rm so}\cdot\sig]$  \cite{revRashba}.
Applied to any spinor, it describes  a rotation of its spin polarization, by an amount which is determined by the distance $s$ and by the spin-orbit ``momentum" $k^{}_{\rm so}$. This rotation is the same for all spinors, and hence does not change the  total  spin polarization. As expected from time-reversal symmetry, the SOI alone cannot generate any spin splitting \cite{bardarson}.
In the presence of ${\bf H}$ the matrix in the square brackets in Eq. (\ref{GG}) is not unitary;  as shown below, this leads to a finite spin polarization, whose magnitude increases with $H$. 

Using these peculiar properties of the spin-dependent tunneling amplitude, we  analyze   the particle current and the magnetization rate.  Both  are determined by  \cite{comsign,sup}
\begin{align}
&R^{L}_{\sigma\sigma'}=\frac{d}{dt}\sum_{\bf k}\langle c^{\dagger}_{{\bf k}\sigma}c^{}_{{\bf k}\sigma'}\rangle\nonumber\\
&=i\sum_{{\bf k},{\bf p},\sigma^{}_{1}}\langle [V^{\ast}_{{\bf k}{\bf p}}]^{}_{\sigma\sigma^{}_{1}} c^{\dagger}_{{\bf p}\sigma^{}_{1}}c^{}_{{\bf k}\sigma'}-
[V^{}_{{\bf k}{\bf p}}]^{}_{\sigma'\sigma^{}_{1}}c^{\dagger}_{{\bf k}\sigma}c^{}_{{\bf p}\sigma^{}_{1}}
\rangle\ ,
\label{RA}
\end{align}
where the angular brackets indicate a quantum average and where we used the assumption that the leads are not polarized.
The total particle current is then
\begin{align}
I^{L}_{}=\sum_{\sigma}R^{L}_{\sigma\sigma}\ , 
\label{IL}
\end{align}
while the magnetization rate, that can be interpreted as a spin current, is
\begin{align}
\dot{\bf M}^{L}_{}=\sum_{\sigma,\sigma'}R^{L}_{\sigma\sigma'}[\sig]^{}_{\sigma\sigma'}\ . 
\label{MR}
\end{align}
The spin-dependent rate Eq. (\ref{RA}) is found to second-order in the tunneling \cite{sup},  
\begin{align}
R^{L}_{\sigma\sigma'}
&=2\pi\sum_{{\bf k},{\bf p}}[V^{}_{\bf kp}V^{\dagger}_{\bf kp}]^{}_{\sigma'\sigma}
\delta(\epsilon^{}_{k}-\epsilon^{}_{p})[f^{}_{L}(\epsilon^{}_{k})-f^{}_{R}(\epsilon^{}_{p})]\ .
\label{Ratef}
\end{align}
This is the Landauer formula, with the $2\times 2$ transmission matrix ${\cal T}=[V^{}_{\bf kp}V^{\dagger}_{\bf kp}]$; it implies that in  the linear-response regime both $I^L_{}$ and $\dot{\bf M}^L_{}$ are proportional to the bias voltage $\mu^{}_L-\mu^{}_R$.  The transmission matrix, ${\cal T}$,  evaluated at the Fermi energy,
is given in terms of the propagator Eq. (\ref{GG}),
${\cal T}=
|C|^2_{}G(d,E^{}_{\rm F})G(d,E^{}_{\rm F})^\dagger_{}$
where $C$ is a constant which is independent of ${\cal H}^{}_{\rm link}$ and $d$ is the length of the wire.  It has the generic form
\begin{align}
{\cal T}^{}_{\sigma'\sigma}={\cal T}^{}_{0}(U\delta^{}_{\sigma'\sigma}+{\bf W}\cdot[\sig]^{}_{\sigma'\sigma})\ ,
\label{T}
\end{align}
where ${\cal T}_{0}$ is the (spin-independent) transmission of the junction in the absence of the external magnetic field and the SOI, and where $U$ and ${\bf W}$ are  a real number and  a real vector, respectively.
For ${\bf H}=0$ the matrix $V^{}_{\bf kp}$ is proportional a unitary matrix \cite{revRashba},  the transmission matrix is proportional to the unit matrix, the rate matrix $R^{L}_{}$ is also proportional to the unit matrix and consequently  $\dot{\bf M}^{L}_{}=0$.
The eigenstates of  this matrix, given by  ${\bf W}\cdot\sig|v^{}_\pm\rangle=\pm W|v^{}_\pm\rangle$, represent spins which are fully polarized in the direction of the vector ${\bf W}$. 

Our central results are the particle current  (i.e., the magnetoconductance) and the magnetization rate (i.e., the spin current),
\begin{align}
I^L_{}\propto {\rm Tr}\{{\cal T}\}
={\cal T}^{}_{0}(2U)\ ,\ 
\dot{\bf M}^{L}_{}\propto {\rm Tr}\{{\cal T}\sig\}
={\cal T}^{}_{0}(2{\bf W})\ .
\label{SF}
\end{align}
When $W=0$ the two eigenvalues of the transmission are identical  and 
the magnetization-rate vector vanishes.  When they  are not equal, $\dot{\bf M}^L_{}$ is directed along the vector ${\bf W}$, and its magnitude is proportional to $W$.

We analyze the charge and spin currents
in two configurations. (i) The external magnetic field  is along the direction of the effective magnetic field due to the SOI,  ${\bf H}^{}_{\parallel}\parallel\hat{\bf h}_{\rm so}$. In this case  $\hat{\bf q}^{}_+=\hat{\bf q}^{}_-=\hat{\bf h}^{}_{\rm so}$, 
$A^{}_{\pm}=k^{}_{\rm F}/[2(k^{2}_{\rm F}+k^{2}_{\rm so}\pm 2m^{\ast}_{}H^{}_{\parallel})^{1/2}]$,
and the  magnetoconductance of the weak link is \cite{comDiv}
\begin{align}
U^{}_{i}=\frac{1+k^{2}_{\rm so}/k^{2}_{\rm F}}{(1+k^{2}_{\rm so}/k^{2}_{\rm F})^{2}-(2m^{\ast}H^{}_{\parallel}/k^{2}_{\rm F})^{2}}\ .
\label{Uparel}
\end{align}
The magnetoconductance increases monotonically with $H^{}_\parallel$ and decreases monotonically with $k^{}_{\rm so}$. 
The magnetization rate  in this configuration 
is directed along ${\bf H}^{}_{\parallel}$, 
\begin{align}
{\bf W}^{}_{i}=
-\frac{2m^{\ast}{\bf H}^{}_{\parallel}/k^{2}_{\rm F}}{(1+k^{2}_{\rm so}/k^{2}_{\rm F})^{2}-(2m^{\ast}H^{}_{\parallel}/k^{2}_{\rm F})^{2}}\ .
\label{Wparel}
\end{align}
The magnetic moment  grows  with $H^{}_{\parallel}$, and  decreases with $k^{}_{\rm so}$. Neither the charge current nor the spin current depend on the length of the weak-link wire.
Since our calculation holds for 
$2m^\ast_{}H^{}_{}< k^{2}_{\rm F}$, the
ratio of the magnetization rate to the particle number rate remains small. 

(ii) The more intriguing configuration is when
${\bf H}\equiv{\bf H}^{}_{\perp}\perp\hat{\bf h}^{}_{\rm so}$
because then the transmission results from the interference of two waves of wave vectors $k_{\pm}$,
\begin{align}
k^{2}_{\pm}=k^{2}_{\rm F}+2k^{2}_{\rm so}\pm 2[
k^{4}_{\rm so}+k^{2}_{\rm so}k^{2}_{\rm F}+(mH^{}_{\perp})^{2}]^{1/2}\ .
\end{align}
These correspond to two spin-projection matrices [see Eqs. (\ref{GG}) and (\ref{qpm})],  determined by the unit vectors
$\hat{\bf q}^{}_{\pm}=[k^{}_{\rm so}k^{}_{\pm}\hat{\bf h}^{}_{\rm so}+m^{\ast}{\bf H}^{}_{\perp}]/[(k^{}_{\rm so}k^{}_{\pm})^{2}+(m^{\ast}H^{}_{\perp})^{2}]^{1/2}$.
In this case the magnetoconductance is given by $U_{ii}$,
\begin{align}
U^{}_{ii}&=2(A^{2}_{+}+A^{2}_{-}+A^{}_{+}A^{}_{-}\cos(\alpha)[1-\hat{\bf q}^{}_{+}\cdot\hat{\bf q}^{}_{-}])\ , 
\label{Uperp}
\end{align}
where $\alpha =(k^{}_{+}-k^{}_{-})d$,  and
\begin{align}
A^{}_{\pm}=2k^{}_{\rm F}
[k^{2}_{\rm so}k^{2}_{\pm}+(m^{\ast}H^{}_{\perp})^{2}]^{1/2}/[k^{}_{\pm}(k^{2}_{+}-k^{2}_{-})]\ ,
\end{align}
The magnetoconductance  oscillates with the length of the weak link.
The oscillations are more pronounced when the difference $U^{}_{ii}-U^{}_{i}$ is plotted for the same magnitudes of the field; this is displayed in Fig. \ref{pU}.

\begin{figure}[htp]
\includegraphics[width=6cm]{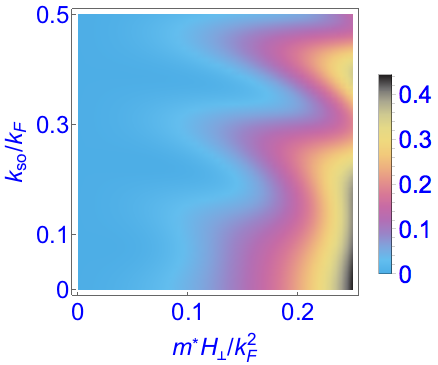}
\caption{(Color online.) The magnetoconductance difference, $U^{}_{ii}-U^{}_{i}$, calculated for $k^{}_{\rm F}d=20$, as function of the spin-orbit coupling ($k^{}_{\rm so}$) measured in units of the Fermi wave vector, and the Zeeman energy, measured in units of $k^{2}_{\rm F}/m^{\ast}$. The oscillations shown are due to the term $\propto \cos(\alpha)$ of $U_{ii}$; $\alpha=(k^{}_+-k^{}_-)d$. 
 }
\label{pU}
\end{figure}

The spin-current  vector ${\bf W}$ in this configuration  is conveniently separated into  components  ${\bf W}_{\rm perp}$ 
along $\hat{\bf q}_{+}\times\hat{\bf q}_{-}\parallel\hat{\bf h}^{}_{\rm so}\times{\bf H}^{}_{\perp}$,    
and ${\bf W}_{\rm plane}$
in the $\{\hat{\bf q}_{+},\hat{\bf q}_{-}\}$ (or $\{\hat{\bf h}^{}_{\rm so},{\bf H}^{}_\perp\})-$plane. 
The magnitudes of those are  \cite{sup}
 \begin{align}
&W^{}_{\rm perp}=2A^{}_{+}A^{}_{-}|\sin(\alpha)\hat{\bf q}^{}_{+}\times\hat{\bf q}^{}_{-}|\ ,
\nonumber\\
&W^{}_{\rm plane}=2\Big |(A^{2}_{+}-A^{2}_{-})^{2}+2A^{}_{+}A^{}_{-}
[1-\hat{\bf q}^{}_{+}\cdot\hat{\bf q}^{}_{-}]\nonumber\\
&\times\Big (A^{}_{+}A^{}_{-}+\cos(\alpha)(A^{2}_{+}+A^{2}_{-})
+A^{}_{+}A^{}_{-}\cos^{2}(\alpha)
\Big )\Big |^{1/2}_{}\ .
\end{align}
The magnitude of ${\bf W}_{\rm plane}$  and
${\bf W}_{\rm perp}$ are displayed in Figs. \ref{Wpar} and \ref{Wperp}, respectively.
The (almost) double periodicity in Fig. \ref{Wperp}  as compared with Fig. \ref{Wpar} appears since 
$|{\bf W}_{\rm perp}|$ is proportional to $\sin(\alpha)$.


\begin{figure}[htp]
\includegraphics[width=6cm]{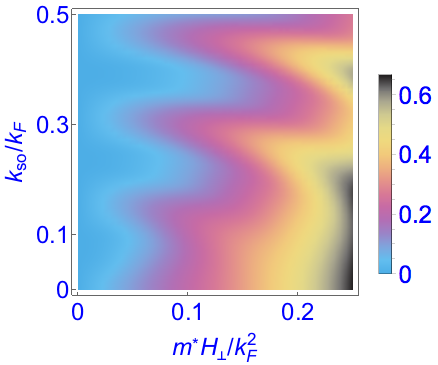}
\caption{(Color online.) The magnitude of the spin current along ${\bf W}_{\rm plane}$  with the same parameters as in Fig. \ref{pU}.
 }
\label{Wpar}
\end{figure}



\begin{figure}[htp]
\includegraphics[width=6cm]{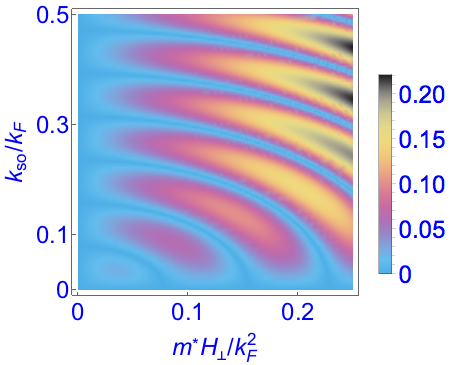}
\caption{(Color online.) The magnitude of the spin current  ${\bf W}_{\rm perp}$  with the same parameters as in Fig. \ref{pU}. 
 }
\label{Wperp}
\end{figure}


In summary, we have shown that  a net amount of charge
and magnetic moment per unit time is transferred through a biased spin-orbit active weak-link from a source to a drain electrode.
Electrons enter and leave the weak-link wire at injection points, whose small volumes are characterized
by the cross-section radius $r^{}_0$ of the wire. Assuming the volumes of the electrodes are much larger than these injection
volumes,  and if the system is part of a closed
electrical circuit, the density of the injected electrons, $\delta n(r)$,  and the density of the associated magnetic moment, $M(r)$,
will decrease with the distance $r$ from the ends of the wire due to a geometrical spreading effect,  
$\delta n(r)/\delta n(r_0)\,, \,\, M(r)/M(r_0) \propto \left( r^{}_0/r\right)^{x}_{}$,
where $x=2$ ($x=1$) in the ballistic (diffusive) transport regime \cite{JETP,LTP}.
The injected magnetization can be measured,  e.g., by
 a properly-positioned 
 SQUID,  or by  a magnetic-resonance force
 microscope.  In an open electrical circuit,  the injection of magnetic moments will lead to
 a finite magnetization, 
 proportional to the bias voltage, 
  in the entire volume of each electrode.

\begin{acknowledgments}

We acknowledge useful discussions with Prof. Junho Suh. OEW and AA are partially supported by the Israel Science Foundation (ISF), by the infrastructure program of Israel Ministry of Science and Technology under contract 3-11173, and by the Pazy Foundation.

\end{acknowledgments}




\begin{thebibliography}{99}
\bibitem{comAB}
This feature distinguishes the present geometry from the devices based on  Aharonov-Bohm interferometers where  the Zeeman energy is compared to the Fermi energy.
\bibitem{compole}
$k_{\pm}^{}$ are real and positive as we assume the Zeeman energy to be smaller than the fermi energy, $H<k^{2}_{\rm F}/(2m^{\ast})$.
\bibitem{comsign}
We consider those two quantities in the left lead; the ones corresponding to the right lead are derived similarly. They have opposite signs to those of the left lead.
\bibitem{comDiv}
The denominators in Eqs. (\ref{Uparel}) and (\ref{Wparel}) vanish when the Zeeman energy approaches the Fermi energy. However, our perturbative calculation requires small tunneling matrix elements, and therefore these equations can be used only when 
 the Zeeman energy  is significantly smaller than the Fermi energy.
\end{thebibliography}
\end{document}


\title{{\underline{Supplemental material for}}\\
{\bf Electric and Magnetic  Gating of Rashba-Active  Weak Links}}

\author{A. Aharony}

\author{O. Entin-Wohlman}

\author{M. Jonson}

\author{R. I. Shekhter
 }
\date{\today}

\maketitle

\noindent {\bf 1. The propagator.} 
The Hamiltonian of an electron moving on  a one-dimensional wire, whose spatial plane wave function is  $\exp[i{\bf k}\cdot {\bf r}]=\exp[iks]$ with the wave vector  ${\bf k}$  directed along the wire, 
is
\begin{align}
{\cal H}=k^2/(2m^\ast_{})-{\bf H}^{}_{\rm eff}({\bf k})\cdot \sig\  , \ \  {\bf H}^{}_{\rm eff}( k)=[k k_{\rm so}/m^{\ast}_{}]\hat{\bf h}^{}_{\rm so}+{\bf H}\ .
\end{align}
The corresponding propagator   is
\begin{align}
G(s;E)&=
\int dk e^{iks}\big[E+i0^+_{}-{\cal H}\big]^{-1}_{}=2m^\ast_{}\int dk e^{iks}\Big[k^2_{\rm F}-k^2+2m^\ast_{}{\bf H}^{}_{\rm eff}(k)\cdot\sig\Big]^{-1}_{}\nonumber\\
\vspace{.5cm}
&=2m^\ast_{}\int dk e^{iks}\frac{k^2_{\rm F}-k^2-2m^\ast_{}{\bf H}^{}_{\rm eff}(k)\cdot\sig}{(k^2_{\rm F}-k^2)^2_{}-(2m^\ast_{}H^{}_{\rm eff}(k))^2_{}}\nonumber\\
\vspace{.5cm}
&=-m^\ast_{}\int dk e^{iks}\Big[\frac{1+\hat{\bf q}(k)\cdot\sig}{k^2_{}-k^2_{\rm F}-2m^\ast_{}H^{}_{\rm eff}(k)}+\frac{1-\hat{\bf q}(k)\cdot\sig}{k^2_{}-k^2_{\rm F}+2m^\ast_{}H^{}_{\rm eff}(k)}\Big]\ .
\label{GGG}
\end{align}
Here $k^{}_{\rm F}$ represents $k^{}_{\rm F}+i0^+_{}$ and 
\begin{align}
\hat{\bf q}(k)={\bf H}^{}_{\rm eff}(k)/H^{}_{\rm eff}(k)\ ,
\end{align}
is a unit vector along the direction of the effective magnetic field. The integrals are calculated by the Cauchy theorem, with $s>0$.
Since $H^{}_{\rm eff}<E^{}_{\rm F}$, the first integrand in the last equality of Eq. (\ref{GGG})  has only one pole in the upper complex $k-$plane, at $k^{}_+=\sqrt{k^{2}_{\rm F}+2m^\ast_{}H^{}_{\rm eff}(k^{}_+)}$. Similarly, the second integrand has only one pole in the upper complex $k-$plane, at $k^{}_-=\sqrt{k^{2}_{\rm F}-2m^\ast_{}H^{}_{\rm eff}(k^{}_-)}$. Both $k^{}_+$ and $k^{}_-$ are roots of a quartic equation. Decomposing the magnetic field ${\bf H}$ into a component ($H_{\parallel}$) along $\hat{\bf h}_{\rm so}$ and a component ($H_{\perp}$) normal to $\hat{\bf h}^{}_{\rm so}$, that equation is
\begin{align}
(k^2-k_{\rm F}^2)^2=[2m^\ast_{}H^{}_{\rm eff}(k)]^2
=(2m^{\ast}_{}H^{}_\perp)^2+(2k^{}_{\rm so}k+2m^{\ast}_{}H^{}_{\parallel})^2\ .
\label{quad}
\end{align}
Performing the integration over a closed contour including (for $s>0$)  semi-circles around each  pole and around the upper complex $k-$plane yields
\begin{align}
G(s;E)&=-i\pi (m^\ast_{}/k^{}_{\rm F})\big[e^{ik^{}_+s}A^{}_+(1+\hat{\bf q}^{}_+\cdot\sig)+e^{ik^{}_-s}A^{}_-(1-\hat{\bf q}^{}_-\cdot\sig)\big]\ ,
\label{res2}
\end{align}
where 
\begin{align}
\hat{\bf q}^{}_\pm\equiv \frac{{\bf H}^{}_{\rm eff}(k^{}_\pm)}{H^{}_{\rm eff}(k^{}_\pm)}=
\frac{k^{}_{\rm so}k^{}_{\pm}\hat{\bf h}^{}_{\rm so}+m^{\ast}{\bf H}}{[(k^{}_{\rm so}k^{}_{\pm}+m^\ast_{}H^{}_\parallel)^{2}+(m^{\ast}H^{}_{\perp})^{2}]^{1/2}}\ ,
\label{qpm}
\end{align}
and $A^{}_\pm$ is the residue of $k^{}_{\rm F}/\big[k^2_{}-k^2_{\rm F}\mp 2m^\ast_{}H^{}_{\rm eff}(k)\big]$ at the poles $k^{}_\pm$ in the first and second integrals in the last equality of Eq. (\ref{GGG}).

When both ${\bf H}$ and $k^{}_{\rm so}$ vanish, $k^{}_\pm=k^{}_{\rm F}$, $A^{}_\pm=1/2$ and $G$ reduces to the free propagator,
\begin{align}
G^{}_0(s;E^{}_{\rm F})=-i\pi m^\ast_{}\exp[ik^{}_{\rm F}s]/k^{}_{\rm F}\ .
\end{align}
Using this expression in  Eq. (\ref{res2}) reproduces the expression for the propagator in the main text,  
\begin{align}
&G(s;E)=G^{}_{0}(s;E)
\big[e^{i\alpha^{}_+}A^{}_+(1+\hat{\bf q}^{}_+\cdot\sig)
+e^{i\alpha^{}_-}A^{}_-(1-\hat{\bf q}^{}_-\cdot\sig)\big]\ ,
\label{GG1}
\end{align}
where $\alpha^{}_{\pm}=(k^{}_{\pm}-k^{}_{\rm F})s$. The propagator is used in the main text to construct the tunneling amplitude of the weak link, whose length is denoted by $d$.

\noindent {\bf 2. Second-order perturbation expansion in the tunneling Hamiltonian.}
As the rate [denoted $R_{\sigma\sigma'}$ in the main text] is proportional to the tunneling amplitude, it suffices   to carry out the perturbation expansion up to first order in the tunneling. This means that we ignore  the effect of the tunneling terms on the reservoirs. Thus
\begin{align}
\langle c^{\dagger}_{{\bf k}\sigma}c^{}_{{\bf p}\sigma'}\rangle &=i\int^{t}dt^{}_{1}\langle {\cal H}^{}_{\rm tun}(t^{}_{1})c^{\dagger}_{{\bf k}\sigma}(t)c^{}_{{\bf p}\sigma'}(t)-c^{\dagger}_{{\bf k}\sigma}(t)c^{}_{{\bf p}\sigma'}(t) {\cal H}^{}_{\rm tun}(t^{}_{1})\rangle
\nonumber\\
&=-i[V^{\ast}_{\bf kp}]^{}_{\sigma\sigma'}\Big (f^{}_{L}(\epsilon^{}_{k})[1-f^{}_{R}(\epsilon^{}_{p})]-
f^{}_{R}(\epsilon^{}_{p})[1-f^{}_{L}(\epsilon^{}_{k})]
\Big )
\int ^{t}dt^{}_{1}e^{i(\epsilon^{}_{k}-\epsilon^{}_{p}+i0^{+}_{})(t-t^{}_{1})}\nonumber\\
&=-[V^{\ast}_{\bf kp}]^{}_{\sigma\sigma'}\frac{
f^{}_{L}(\epsilon^{}_{k})-
f^{}_{R}(\epsilon^{}_{p})}{
\epsilon^{}_{k}-\epsilon^{}_{p}+i0^{+}_{}}
\ ,
\end{align}
and
\begin{align}
\langle c^{\dagger}_{{\bf p}\sigma}c^{}_{{\bf k}\sigma'}\rangle &=i\int^{t}dt^{}_{1}\langle {\cal H}^{}_{\rm tun}(t^{}_{1})c^{\dagger}_{{\bf p}\sigma}(t)c^{}_{{\bf k}\sigma'}(t)-c^{\dagger}_{{\bf p}\sigma}(t)c^{}_{{\bf k}\sigma'}(t) {\cal H}^{}_{\rm tun}(t^{}_{1})\rangle
\nonumber\\
&=i[V^{}_{\bf kp}]^{}_{\sigma'\sigma}\Big (f^{}_{L}(\epsilon^{}_{k})[1-f^{}_{R}(\epsilon^{}_{p})]-
f^{}_{R}(\epsilon^{}_{p})[1-f^{}_{L}(\epsilon^{}_{k})]
\Big )
\int^{t} dt^{}_{1}e^{i(\epsilon^{}_{p}-\epsilon^{}_{k}+i0^{+}_{})(t-t^{}_{1})}\nonumber\\
&=[V^{}_{\bf kp}]^{}_{\sigma'\sigma}\frac{
f^{}_{L}(\epsilon^{}_{k})-
f^{}_{R}(\epsilon^{}_{p})}{
\epsilon^{}_{p}-\epsilon^{}_{k}+i0^{+}_{}}
\ .
\end{align}
It follows that
\begin{align}
R^{L}_{\sigma\sigma'}&=i\sum_{{\bf k},{\bf p},\sigma^{}_{1}}\Big ([V^{\ast}_{{\bf k}{\bf p}}]^{}_{\sigma\sigma^{}_{1}}
[V^{}_{\bf kp}]^{}_{\sigma'\sigma^{}_{1}}\frac{
f^{}_{L}(\epsilon^{}_{k})-
f^{}_{R}(\epsilon^{}_{p})}{
\epsilon^{}_{p}-\epsilon^{}_{k}+i0^{+}_{}}+[V^{}_{{\bf k}{\bf p}}]^{}_{\sigma'\sigma^{}_{1}}[V^{\ast}_{\bf kp}]^{}_{\sigma\sigma^{}_{1}}\frac{
f^{}_{L}(\epsilon^{}_{k})-
f^{}_{R}(\epsilon^{}_{p})}{
\epsilon^{}_{k}-\epsilon^{}_{p}+i0^{+}_{}}\Big )\nonumber\\
&=2\pi\sum_{{\bf k},{\bf p}}[V^{}_{\bf kp}V^{\dagger}_{\bf kp}]^{}_{\sigma'\sigma}
\delta(\epsilon^{}_{k}-\epsilon^{}_{p})[f^{}_{L}(\epsilon^{}_{k})-f^{}_{R}(\epsilon^{}_{p})]\ .
\label{Rate1}
\end{align}

\noindent{\bf
3. The transmission matrix.}
The transmission matrix is
\begin{align}
[V^{}_{\bf kp}V^{\dagger}_{\bf kp}]^{}_{\sigma'\sigma}={\cal T}^{}_{0}[U\delta^{}_{\sigma'\sigma}+{\bf W}\cdot[\sig]^{}_{\sigma'\sigma}]\ ,
\label{T}
\end{align}
where ${\cal T}_{0}$ is the transmission in the absence of the spin-orbit coupling and the external magnetic field.
Using Eq. (\ref{res2}), the spin-dependent part of the amplitude product (\ref{T}) is
\begin{align}
&U+{\bf W}\cdot\sig=\nonumber\\
&\Big (e^{i\alpha^{}_{+}}A^{}_{+}(1+\sig\cdot\hat{\bf q}^{}_{+})+e^{i\alpha^{}_{-}}A^{}_{-}(1
-\sig\cdot\hat{\bf q}^{}_{-})\Big )\Big (e^{-i\alpha^{}_{+}}A^{}_{+}(1+\sig\cdot\hat{\bf q}^{}_{+})+e^{-i\alpha^{}_{-}}A^{}_{-}(1
-\sig\cdot\hat{\bf q}^{}_{-})\Big )\nonumber\\
&=2(A^{2}_{+}[1+\sig\cdot{\bf q}^{}_{+}]+A^{2}_{-}[1-\sig\cdot{\bf q}^{}_{-}])+2A^{}_{+}A^{}_{-}\cos(\alpha)\Big [1-\hat{\bf q}^{}_{+}\cdot\hat{\bf q}^{}_{-}+\sig\cdot (\hat{\bf q}^{}_{+}-\hat{\bf q}^{}_{-})\Big ]\nonumber\\
&\hspace{3cm}+2A^{}_{+}A^{}_{-}\sin(\alpha)\sig\cdot\hat{\bf q}^{}_{+}\times\hat{\bf q}^{}_{-}\ ,
\label{UW}
\end{align}
where
$\alpha=\alpha^{}_+-\alpha^{}_-$.

\noindent
{\bf 4. Magnetic field parallel to the effective magnetic field of the spin-orbit interaction.}
In this case, 
\begin{align}
k^2_{}-k^2_{\rm F}\mp 2m^\ast_{}H^{}_{\rm eff}(k)=k^2_{}-k^2_{\rm F}\mp 2(k^{}_{\rm so}k+m^\ast_{}H^{}_\parallel)=0\ ,
\end{align}
with the positive roots
\begin{align}
k^{}_{\pm}&=\sqrt{k^{2}_{\rm F}\pm 2mH^{}_{\parallel}+k_{\rm so}^{2}}\pm k_{\rm so}^{}
\end{align}
and the corresponding negative roots $-\sqrt{k^{2}_{\rm F}\pm 2mH^{}_{\parallel}+k_{\rm so}^{2}}\pm k_{\rm so}^{}$.
It follows that
\begin{align}
A^{}_{\pm}=\frac{k^{}_{\rm F}}{2\sqrt{k^2_{\rm F}+ k^{2}_{\rm so}\pm2m^\ast_{}H^{}_\parallel}}\ ,
\end{align}
and therefore
\begin{align}
U^{}_{i}=\frac{1+k^{2}_{\rm so}/k^{2}_{\rm F}}{(1+k^{2}_{\rm so}/k^{2}_{\rm F})^{2}-(2m^{\ast}H^{}_{\parallel}/k^{2}_{\rm F})^{2}}
\label{Uparel}
\end{align}
and
\begin{align}
{\bf W}^{}_{i}=2(A^2_+-A^2_-)\hat{\bf h}^{}_{\rm so}=
-\frac{2m^{\ast}{\bf H}/k^{2}_{\rm F}}{(1+k^{2}_{\rm so}/k^{2}_{\rm F})^{2}-(2m^{\ast}H^{}_{\parallel}/k^{2}_{\rm F})^{2}}\ ,
\label{Wparel}
\end{align}
as given in the main text.

\noindent{\bf 5. Magnetic field normal to the effective magnetic field of the spin-orbit interaction.}
In this case Eq. (\ref{quad}) becomes
\begin{align}
(k^2-k_F^2)^2=(2m^\ast_{}H^{}_\perp)^2+(2k^{}_{\rm so}k)^2\ ,
\end{align}
and the relevant (positive) poles are given by
\begin{align}
k^{2}_{\pm}=2k^{2}_{\rm so}+k^{2}_{\rm F}\pm 2\sqrt{k^{4}_{\rm so}+k^{2}_{\rm so}k^{2}_{\rm F}+(mH^{}_{\perp})^{2}}\ .
\end{align}
The denominator in the second line in Eq. (\ref{GGG}) is then
$(k-k^{}_+)(k-k^{}_-)(k+k^{}_+)(k+k^{}_-)$, and therefore 
\begin{align}
 A^{}_{\pm}=2k^{}_{\rm F}
\frac{[(k^{}_{\rm so}k^{}_{\pm})^{2}+(m^{\ast}_{}H^{}_{\perp})^{2}]^{1/2}_{}}{k^{}_{\pm}(k^{2}_{+}-k^{2}_{-})}
\ ,\label{A}
\end{align}
and
\begin{align}
\hat{\bf q}^{}_{\pm}=\frac{k^{}_{\rm so}k^{}_{\pm}\hat{\bf h}^{}_{\rm so}+m^{\ast}_{}{\bf H}^{}_{\perp}}{
\sqrt{(k^{}_{\rm so}k^{}_{\pm})^{2}+(m^{\ast}_{}H^{}_{\perp})^{2}}}\ .
\label{qpm}
\end{align}
Using Eqs. (\ref{A}) and (\ref{qpm}), one finds
\begin{align}
1+\hat{\bf q}^{}_{+}\cdot\hat{\bf q}^{}_{-}&=
\frac{k^{2}_{\rm so}k^{2}_{\rm F}[1+\sqrt{1-(2m^{\ast}H^{}_{\perp}/k^{2}_{\rm F})^{2}}]+2(m^{\ast}_{}H^{}_{\perp})^{2}}{k^{2}_{\rm so}k^{2}_{\rm F}+(m^{\ast}_{}H^{}_{\perp})^{2}}\ ,\nonumber\\
1-\hat{\bf q}^{}_{+}\cdot\hat{\bf q}^{}_{-}&=
\frac{k^{2}_{\rm so}k^{2}_{\rm F}[1-\sqrt{1-(2m^{\ast}H^{}_{\perp}/k^{2}_{\rm F})^{2}}]}
{k^{2}_{\rm so}k^{2}_{\rm F}+(m^{\ast}_{}H^{}_{\perp})^{2}}
\ ,
\label{qq}
\end{align}
and
\begin{align}
A^{2}_{+}+A^{2}_{-}
&=\frac{k^2_{\rm F}\big[k^2_{\rm so}k^4_{\rm F}+(m^\ast_{}H^{}_\perp)^2_{}(k^2_{\rm F}-2k^2_{\rm so}\big]}{2[k^4_{\rm F}-(2m^\ast_{}H^{}_\perp)^2_{}][k^4_{\rm so}+k^2_{\rm so}k^2_{\rm F}+(m^\ast_{}H^{}_\perp)^2]}
\ ,\nonumber\\
A^{2}_{+}-A^{2}_{-}&=\frac{1}{1-(2m^{\ast}H^{}_{\perp}/k^{2}_{\rm F})^{2}}\times\frac{-(m^{\ast}_{}H^{}_{\perp})^{2}/k^{2}_{\rm F}}{\sqrt{k^{4}_{\rm so}+k^{2}_{\rm so}k^{2}_{\rm F}+(m^{\ast}_{}H^{}_{\perp})^{2}}}
\ ,\nonumber\\
A^{}_{+}A^{}_{-}&=\frac{1}{\sqrt{1-(2m^{\ast}H^{}_{\perp}/k^{2}_{\rm F})^{2}}}\times\frac{k^{2}_{\rm so}k^{2}_{\rm F}+(m^{\ast}_{}H^{}_{\perp})^{2}}{4k^{4}_{\rm so}+4k^{2}_{\rm so}k^{2}_{\rm F}+(2m^{\ast}_{}H^{}_{\perp})^{2}}
\ .
\label{AA}
\end{align}
Therefore,
\begin{align}
U^{}_{ii}&=2\Big (\frac{1}{1-(2m^{\ast}H^{}_{\perp}/k^{2}_{\rm F})^{2}}\times\frac{2(m^{\ast}_{}H^{}_{\perp})^{2}+2k^{2}_{\rm so}k^{2}_{\rm F}[1-(2m^{\ast}_{}H^{}_{\perp})^{2}/(2k^{4}_{\rm F})]}{4k^{4}_{\rm so}+4k^{2}_{\rm so}k^{2}_{\rm F}+(2m^{\ast}_{}H^{}_{\perp})^{2}}\nonumber\\
&+
\frac{1}{\sqrt{1-(2m^{\ast}H^{}_{\perp}/k^{2}_{\rm F})^{2}}}\times\frac{k^{2}_{\rm so}k^{2}_{\rm F}[1-\sqrt{1-(2m^{\ast}H^{}_{\perp})^{2}/k^{4}_{\rm F}}]}{4k^{4}_{\rm so}+4k^{2}_{\rm so}k^{2}_{\rm F}+(2m^{\ast}_{}H^{}_{\perp})^{2}}\cos(\alpha)\Big )\ .
\end{align}

The  vector
${\bf W}_{\rm perp}$ (the component of  ${\bf W}$ normal to the plane spanned by $\hat{\bf q}^{}_{+}$  and $\hat{\bf q}^{}_{-}$) is proportional to $\sin(\alpha)$, with the amplitude
\begin{align}
&2A^{}_{+}A^{}_{-}\sqrt{1-(\hat{\bf q}^{}_{+}\cdot\hat{\bf q}^{}_{-})^{2}}\nonumber\\
&=\frac{2\sqrt{2}(m^{\ast}H^{}_\perp)k^{}_{\rm so}}{\sqrt{1-(2m^{\ast}H^{}_{\perp}/k^{2}_{F})^{2}}}\times\frac{\Big [2k^{2}_{\rm so}+k^{2}_{\rm F}-\sqrt{k^{4}_{\rm F}-(2m^{\ast}H^{}_{\perp})^{2}}\Big ]^{1/2}}{4k^{4}_{\rm so}+4k^{2}_{\rm so}k^{2}_{\rm F}+(2m^{\ast}_{}H^{}_{\perp})^{2}}\nonumber\\
&=
\frac{2\sqrt{2}(m^{\ast}H^{}_\perp)k^{}_{\rm so}}{\sqrt{1-(2m^{\ast}H^{}_{\perp}/k^{2}_{\rm F})^{2}}}\times\frac{\Big [2k^{2}_{\rm so}+k^{2}_{\rm F}+\sqrt{k^{4}_{\rm F}-(2m^{\ast}H^{}_{\perp})^{2}}\Big ]^{-1/2}}{\Big [4k^{4}_{\rm so}+4k^{2}_{\rm so}k^{2}_{\rm F}+(2m^{\ast}_{}H^{}_{\perp})^{2}\Big ]^{1/2}}\ .
\end{align}
The magnitude squared of the vector ${\bf W}_{\rm plane}$ (the component of  ${\bf W}$ in the plane spanned by $\hat{\bf q}^{}_{+}$  and $\hat{\bf q}^{}_{-}$) is 
\begin{align}
|{\bf W}^{}_{\rm plane}|^{2}&=4\Big (A^{4}_{+}+A^{4}_{-}-2A^{2}_{+}A^{2}_{-}\hat{\bf q}^{}_{+}\cdot\hat{\bf q}^{}_{-}
\nonumber\\
&+2A^{}_{+}A^{}_{-}[1-\hat{\bf q}^{}_{+}\cdot\hat{\bf q}_{-}]\cos(\alpha)[A^{2}_{+}+A^{2}_{-}+A^{}_{+}A^{}_{-}\cos(\alpha)]\Big )\nonumber\\
&=4\Big [(A^{2}_{+}-A^{2}_{-})^{2}\nonumber\\
&+2A^{}_{+}A^{}_{-}[1-\hat{\bf q}^{}_{+}\cdot\hat{\bf q}_{-}]
\Big (A^{}_{+}A^{}_{-}+\cos(\alpha)[A^{2}_{+}+A^{2}_{-}]+\cos^{2}(\alpha)A^{}_{+}A^{}_{-}\Big )\Big ]\ .
\end{align}
Inserting here Eqs. (\ref{qq}) and (\ref{AA}) gives
\begin{align}
|{\bf W}^{}_{\rm plane}|^{2}&=4\Big [
\frac{1}{[1-(2m^{\ast}H^{}_{\perp}/k^{2}_{\rm F})^{2}]^{2}}\times\frac{(m^{\ast}_{}H^{}_{\perp})^{4}/k^{4}_{\rm F}}{k^{4}_{\rm so}+k^{2}_{\rm so}k^{2}_{\rm F}+(m^{\ast}_{}H^{}_{\perp})^{2}}\nonumber\\
&+\frac{1-\sqrt{1-(2m^{\ast}_{}H^{}_{\perp}/k^{2}_{\rm F})^{2}}}{
\sqrt{1-(2m^{\ast}_{}H^{}_{\perp}/k^{2}_{\rm F})^{2}}}
\frac{2k^{2}_{\rm so}k^{2}_{\rm F}}{4k^{4}_{\rm so}+4k^{2}_{\rm so}k^{2}_{\rm F}+(2m^{\ast}_{}H^{}_{\perp})^{2}}\nonumber\\
&\times\Big (\frac{1}{\sqrt{1-(2m^{\ast}H^{}_{\perp}/k^{2}_{\rm F})^{2}}}\times\frac{k^{2}_{\rm so}k^{2}_{\rm F}+(m^{\ast}_{}H^{}_{\perp})^{2}}{4k^{4}_{\rm so}+4k^{2}_{\rm so}k^{2}_{\rm F}+(2m^{\ast}_{}H^{}_{\perp})^{2}}[1+\cos^{2}(\alpha)]\nonumber\\
&+\frac{1}{1-(2m^{\ast}H^{}_{\perp}/k^{2}_{\rm F})^{2}}\times\frac{2(m^{\ast}_{}H^{}_{\perp})^{2}+2k^{2}_{\rm so}k^{2}_{\rm F}[1-(2m^{\ast}_{}H^{}_{\perp})^{2}/(2k^{4}_{\rm F})]}{4k^{4}_{\rm so}+4k^{2}_{\rm so}k^{2}_{\rm F}+(2m^{\ast}_{}H^{}_{\perp})^{2}}\cos(\alpha)\Big )\Big ]\ .
\end{align}